\documentclass{llncs}
\usepackage[utf8]{inputenc}
\usepackage{url}
\usepackage{booktabs}
\usepackage{marvosym}
\usepackage{graphicx}
\usepackage{amssymb}

\newcommand{\todo}[1]{}

\begin{document}

\mainmatter


\title{Recommending Scientific Literature: Comparing Use-Cases and Algorithms}
 

\author{Roman Kern\inst{1} \and Kris Jack\inst{2} \and Michael Granitzer\inst{3}}
\institute{Institute of Knowledge Management, Graz University of Technology\\
\url{rkern@tugraz.at} \\ \and
Mendeley Ltd.\\
\url{kris.jack@mendeley.com} \\ \and
University of Passau \\
\url{michael.granitzer@uni-passau.de}
}

\maketitle
\begin{abstract}

An important aspect of a researcher's activities is to find relevant and related publications.
The task of a recommender system for scientific publications is to provide a list of papers that match these criteria.
Based on the collection of publications managed by Mendeley, four data sets have been assembled that reflect different aspects of relatedness.
Each of these relatedness scenarios reflect a user's search strategy.
These scenarios are public groups, venues, author publications and user libraries.
The first three of these data sets are being made publicly available for other researchers to compare algorithms against.
Three recommender systems have been implemented: a collaborative filtering system; a content-based filtering system; and a hybrid of these two systems.
Results from testing demonstrate that collaborative filtering slightly outperforms the content-based approach, but fails in some scenarios.
The hybrid system, that combines the two recommendation methods, provides the best performance, achieving a precision of up to 70\%.
This suggests that both techniques contribute complementary information in the context of recommending scientific literature and different approaches suite for different information needs.

\end{abstract}

%
%

\section{Introduction}

Researchers need to keep up-to-date with the latest publications in their field.
As the number of articles published steadily increases, the role that information retrieval systems plays becomes increasingly important.
A recommender system that is capable of automatically leading users to related research, based on their current interests, would likely save them time and help them to perform a more thorough background search.
In order to build such a system, the main use-cases for such a task need to be identified.
There cannot be only one criterion for relatedness between publications, as researchers may look for different kinds of publications, depending on their current information needs and current context.

\begin{figure}[t!]
\begin{center}
\includegraphics[width=.5\columnwidth]{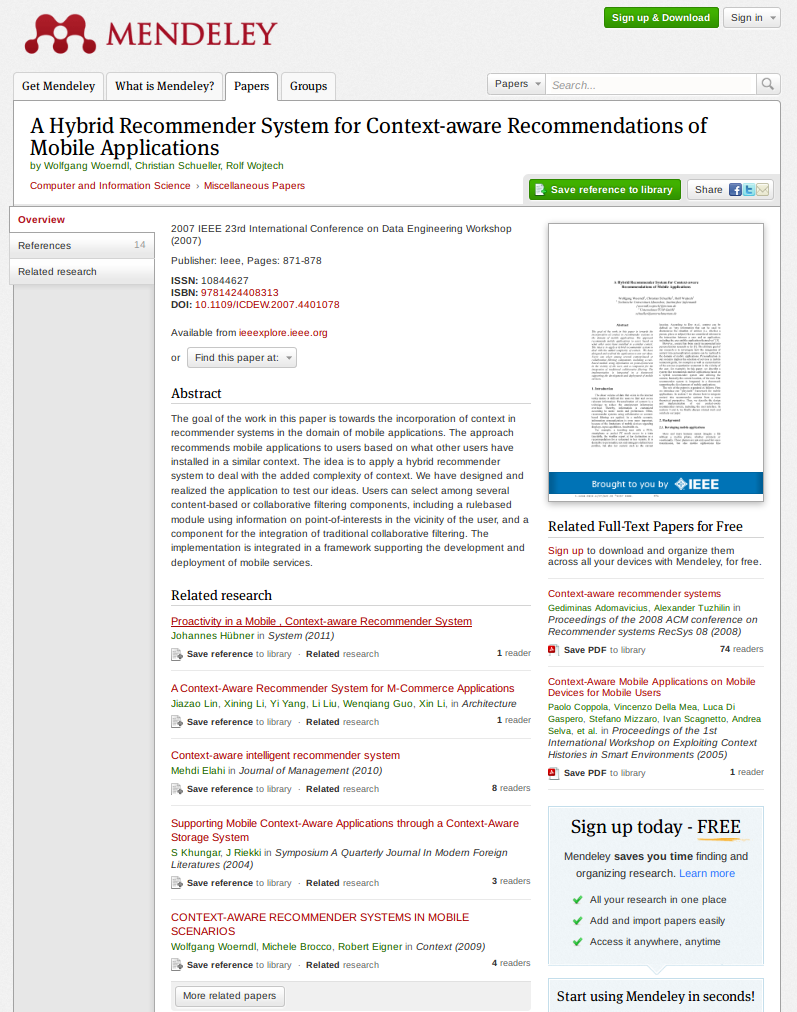}
\end{center}
\caption[Example Mendeley Article Page]{
A scientific article as displayed on Mendeley's web site.
It makes reference to five other related articles, which should help the researcher to contextualise the main article and discover related research.
}
\label{fig:mendeley-article-page}
\end{figure}

Mendeley's database of scientific literature \cite{Henning2008} serves as the basis for evaluating the recommender systems presented here.
Mendeley already attempts to generate related research automatically using a content-based filtering algorithm (Figure~\ref{fig:mendeley-article-page}).
It is similar to one of the algorithms presented in this paper.
This database has been crowdsourced from almost two million researchers \cite{Hammerton2012} who use Mendeley's tools to organise their research, collaborate with others and discovery new research.
Based on the data available in the Mendeley database, distinct use-cases were identified, that reflect different common scenarios.
From these scenarios, four data sets have been constructed.

The decisions behind constructing the data sets and developing an evaluation procedure were guided not only to provide recommender systems that work according to the data sets. 
The aim of this study is also to gain a better understanding of how people organise their scientific literature.
Thus the contributions of this paper can be summarised as:
I) Create real-world data sets that represent different aspect of relatedness of academic articles,
II) Develop and evaluate two recommender systems based on state-of-the-art techniques,
III) Propose a hybrid system that combines collaborative and content based filtering,
IV) Analysis of the systems performance to draw conclusions about the users behaviour.


\section{Related Work}

Recommendation system technology has a huge potential for supporting researchers in keeping up-to-date with research and making new discoveries.  
Perhaps surprising then, the uptake of recommender services has been slow in scientific digital libraries.  
\cite{Franke2008} argues that difficulties in mechanism design and decreased funding for scientific libraries can partially explain this deficiency.  
The growing popularity of services such as CiteSeer, Mendeley and PubMed, as of late, however, has seen an increased use of recommenders in the scientific domain.
\cite{Beel2010} investigated the problem of constructing appropriate data sets for evaluating research paper relatedness.  
They tested the hypothesis that papers appearing together in a single mind map, created by researchers when drafting papers, tend to be more related to one another than papers that do not appear in the same mind map. 
This intuitively appealing idea was put to the test in a small scale experiment with five participants.  
Each participant was shown pairs of research articles and asked to rate their relatedness on a five-point Likert scale.  
The results of the study supported the hypothesis that papers appearing in the same mind map are judged as being more related to one another than papers that do not.  
Mendeley Groups, that allow Mendeley users to connect researchers and research papers together, are similar to such mind maps, suggesting that they too could be an interesting source of research paper relatedness.

Scienstein offers a suite of tools to support researchers in managing their libraries.  
\cite{Gipp2009a} describe their algorithm for retrieving research papers that are related to one another.  
Primarily, they exploit citation graphs, looking at backward and forward citations for papers.  
%
While citation networks provide strong cues to article relatedness, it can take several years for an article to be appropriately embedded in a network due to the time that it takes to publish articles.  
\cite{Pohl2007} investigated the use of real-time article click data generated by logging user actions in digital scientific libraries as a cue for article relatedness.  

PubMed generates related research for articles in their catalogue.  
They trained a probabilistic topic-based model to learn article relatedness based on article meta-data (e.g. title, abstract and MeSH terms).  
In testing on the TREC 2005 genomics track (including MEDLINE abstracts), the probabilistic topic-based model is shown to slightly outperform BM25, with 0.399 vs. 0.383 precision at 5, respectively \cite{Lin2007a}.  
By taking a random week long sample of logs, PubMed show that out of two million user sessions, 19\% of them include at least one click on an article link provided as related research.


\section{Testing Data}



Four testing data sets have been created based on these four types of relatedness, the first three of them being made publicly available for non-commercial research purposes\footnote{\url{http://team-project.tugraz.at/the-project/results/}}.  
These testing sets provide ground truth benchmarks in order to test the performance of recommenders in retrieving related research.
These new data sets accompanying an already significant list of data that Mendeley provides for researchers\cite{Jack2012b}.

In order to simplify the number of variables in the experiments, all of the testing data sets share similar characteristics.  
The testing data sets contain multiple sets of articles.  
Each article set can be seen to represent a topic, where articles that appear in the same set are considered to be related to one another.  
Article set membership is mutually exclusive within a testing set.  
That is, no article can appear in more than one article set within a single testing set.  
By excluding overlaps between sets of related documents, it is expected to simplify interpretation of the experimental results.  
Second, all article sets have a fixed cardinality of between 10 and 20 articles.  This makes for rather small and constrained topics.

\textbf{Public Groups} Mendeley groups bring together articles and researchers for a diverse range of use cases from performing a collaborative literature review to generating a reading list for a university course.
In many cases, it is has been observed that the articles appearing in public groups tend to share a common topic.  

\textbf{Venues} The venues data set is the largest of the four, with almost 40,000 article sets.
These sets were constructed by grouping articles together from Mendeley's catalogue by the venue in which they were published.
This set reflects the use-case of a research looking for related papers based on the publication venue.

\textbf{Author Publications} The author publications testing set is the smallest of the four data sets with a total of 2,654 article sets.
This data set has been constructed by taking the articles that Mendeley users claim to have published themselves by adding them to their ``My Publications'' folder in Mendeley Desktop.  
The use-case for this data-set is a researcher looking for more papers from the same author.

\textbf{User Libraries} The final testing set is constructed from the contents of Mendeley user libraries.
This testing set is also quite large, with over 11,000 article sets.
It is a random sample of a small percentage of Mendeley user libraries that have between 10 and 20 articles in them.  



\section{Experimental Setup}


\subsection{Procedure}

A query article is randomly selected from an article set in a test set.  
The task of the algorithm is to retrieve the other articles that appear in the same article set as the query article (i.e. the related documents).  
The search space in which the algorithm performs the retrieval includes all articles that appear in the given test set, excluding the query article.  
Due to data set construction constraints, an article cannot appear more than once in a given test set.  
This procedure is followed twice for each article set, selecting two articles at random, and repeated for every article set that appears in a test set.


Three training data sets have been constructed for use in the four test scenarios.
The main training data set is made up of Mendeley user libraries that have between 2 and 1,000 articles.  
Libraries with just one article are excluded since they do not contain useful information and libraries with more than 1,000 articles are excluded since the article relationships provide little discriminative value.  
The training set contains over 325,000 user libraries.  
In the first two testing scenarios, where the model aims to retrieve related articles based on Mendeley groups and publication venues, this training data set is used in full.  
In the third scenario of predicting relatedness based on author publications, however, all authors of the publications in the testing set are removed from the training data set.  
Similarly, for the fourth scenario where relatedness is based on user libraries, all of these user libraries are removed from the training set.  
The same training data sets are used to train the content-based filtering, collaborative filtering and hybrid systems.

\subsection{Algorithms}

Recommender systems can typically be categorised into three different families \cite{Candillier2009}: content-based filtering; collaborative filtering; and a hybrid.
In this study, the first approach of content-based filtering uses the meta-data of the articles.
The second is trained on articles readership, based on Mendeley user libraries.
The third takes the output of the collaborative filtering component and integrates it into the content-based algorithm, producing a hybrid model.

\subsubsection{Content Based Filtering}

The content-based filtering algorithm exploits the available meta-data for articles.
Therefore an important aspect of this approach is the different level of sparseness of the individual meta-data types.
While the title is present for each article, other kinds of meta-data, for example tags, is only present for a small subset. 


To study the impact of different text processing approaches, a series of evaluation were conducted on a preliminary data set.
A number of tokenisation strategies were employed: stemming; stop-word removal; character n-grams; and word bi-grams.
In preliminary tests, the best performance was achieved when indexing words as uni-grams, while stemming had little impact.

The algorithm is implemented using Apache Lucene\footnote{\url{http://lucene.apache.org/core/} (Version 3.3)}, an open-source search library.
Lucene allows each indexed document to consist of multiple, individual fields, which can be queried separately or in combinations.
Each article meta-data type is indexed as a separate field.
For each of the four data sets, a separate search index was constructed, where each index consists of documents representing articles, with a set of fields, one for each meta-data type.
With the exception of the title and abstract fields, no further text processing has been applied.

The meta-data fields were a) \textbf{Title}: The title split into words, which are then normalised to lower-case. 
    The terms in the title are stemmed with the Porter stemmer and the default stop-word filtering of Lucene is applied.
b) \textbf{Abstract}: The abstract is split into words, normalised to lower-case, but not stemmed.
c) \textbf{Author}: Each author name of an article is processed as a single term.
d) \textbf{Tag}: The tags assigned by Mendeley users to articles.
    A tag must have been applied by three different users, across Mendeley's entire user base, in order to appear in an article's meta-data.
e) \textbf{Keyword}: The keywords as supplied by the authors of the articles.
f) \textbf{Mesh-Term}: Keywords from the controlled vocabulary thesaurus used by PubMed.
    MeSH terms show a high overlap with the author supplied keywords.
g) \textbf{Textrank Keyword}: Keywords automatically extracted from the abstract by applying the TextRank\cite{Mihalcea2004} algorithm.
    These terms are only present for article where an abstract is present.
Other types of meta-data were also indexed such as the article's disciplines and year, but a detailed description is skipped here for brevity, as they are not used in the reported numbers of the evaluation section.
For the content-based recommender approach, two mainstream algorithms, TFIDF\cite{Salton1988} and BM25\cite{Robertson1996}, were implemented.
The first three configurations are built upon the searching and scoring functionalities provided by Lucene.
%
%
%
%
%
%

To improve the performance of the content base recommender, techniques from the field of information retrieval were adapted. 
These were then realised in a number of configurations: 
i) \textbf{Configuration TFIDF:} Build a single query for all meta-data types, 
ii) \textbf{Configuration NQ:} Introduces a discounting factor for pairs of articles, which share just a few meta-data fields,
iii) \textbf{Configuration PRF:} The third configuration implements a form of pseudo relevance feedback.
This technique is also known as blind feedback and can be seen as local query expansion\cite{Xu1996}.
%
%
iv) \textbf{Configuration BM25C:} The fourth and final configuration does not employ the Lucene retrieval facilities, although it uses the same text processing methods.
Instead, here the BM25 retrieval function\cite{Robertson1996} is used as a term weighting scheme.
%
%
%
%

\subsubsection{Collaborative Filtering}

The second algorithm follows an item-based collaborative filtering heuristic.
Apache Mahout's\footnote{\url{http://mahout.apache.org/} (Version 0.6)} implementation of an item-based similarity job is used. 
No modifications have been made to the out-of-the-box version of this implementation making the results from the experiments conducted here more easily reproducible.
Mahout has been selected for its scaling properties, being horizontally scalable over Apache's distributed system, Hadoop\footnote{\url{http://hadoop.apache.org/}}, easily coping with data sets that contain hundreds of millions of relationships.

Two of the item-based collaborative filtering parameters are varied in the evaluation, namely the similarity measure and maximum number of similarities considered per item, while the others retain their default values: maximum number of cooccurrences considered per item (100); and only boolean data is provided (e.g. \emph{user x has article y}, with no strength of preference indicated).

\subsubsection{Hybrid System}

The third algorithm implements a hybrid solution that combines the information from both the content-based and collaborative filtering systems.  
This algorithm, similar to the content-based one, is implemented using Lucene.
The output of the collaborative filtering algorithm for each of the articles in the test data set is treated as another source of meta-data that describes the articles.
Instead of words (e.g. an article title or abstract), the internal unique ids of the articles are mapped as terms when building the search index.
By following this approach, the output of collaborative filtering can be used in all four configurations of the content-based filtering method.
As the BM25 retrieval function is tailored towards natural language text, the output of the collaborative filtering is not weighted for the BM25C configuration.

\section{Results}



\begin{table}[t!]
\caption{Performance of the collaborative filtering recommender system for the four data sets.}
\label{tbl:cf-results}
\begin{center}
    \begin{tabular}{ lrrrr }
    \toprule
    \textbf{Measure} & \textbf{Groups} & \textbf{Venues} & \textbf{Pubs.} & \textbf{Libs.} \\ \midrule
    P@5 & 0.407 & 0.049 & 0.460 & 0.039 \\
    \bottomrule
    \end{tabular}
\end{center}
\end{table}


\texttt{trec\_eval}\footnote{\url{http://trec.nist.gov/trec_eval/}} has been used to evaluate the results from testing.
Given two files, a ground truth file and a file generated by a recommender system, the tool computes a number of metrics.
In the following, the performance of the various algorithms is reported in terms of the precision at five measure (P@5), which is appropriate for Mendeley's use case of displaying five related articles (Figure~\ref{fig:mendeley-article-page}).

\textbf{Collaborative Filtering} The collaborative filtering approach appears to works particularly well for the data sets ``groups'' and ``publications'' (Table~\ref{tbl:cf-results}). 
For author publications, about half of all top five recommendations are correct according to the ground truth.
The measured precision for the other two data sets is rather low.
Six similarity metrics have also been tested in order to investigate which one performs best in the groups scenario (Table~\ref{tbl:cf-sim-measures}).

\begin{table}[t!]
\caption{Performance of the collaborative filtering recommender system for different similarity measures for the public groups data set.}
\label{tbl:cf-sim-measures}
\begin{center}
    \begin{tabular}{ rrrrrr }
    \toprule
    \textbf{City} & \textbf{Cooc} & \textbf{Cosine} & \textbf{LogL} & \textbf{Tani} & \textbf{L2} \\ \midrule
    0.330 & \textbf{0.407} & 0.366 & \textbf{0.407} & 0.399 & 0.330 \\
    \bottomrule
    \end{tabular}
\end{center}
\end{table}

\textbf{Content-based Filtering} The content-based approach performs differently depending upon which meta-data field is used and the scenario under testing (Table~\ref{tbl:cb-tfidf-results}).
Using a single field, titles tend to produce the best results, while combinations of fields outperform single fields alone.
Precision for the ``groups'' and ``publications'' scenarios tends to be higher than for the other two.

\begin{table}[t!]
\caption{Performance of the content-based filtering recommender system using the basic TFIDF configuration, measured via P@5. Runs that contain the ground truth are marked with a $^{\dag}$ symbol.}
\label{tbl:cb-tfidf-results}
\begin{center}
    \begin{tabular}{ lrrrr }
    \toprule
    \textbf{Metadata} & \textbf{Groups} & \textbf{Venues} & \textbf{Pubs.} & \textbf{Libs.} \\ \midrule
    Title & 0.277 & 0.092 & 0.299 & 0.183 \\
    Abstract & 0.170 & 0.024 & 0.260 & 0.091 \\
    Tag & 0.014 & 0.001 & 0.003 & 0.002 \\
    Author & 0.121 & 0.094 & 0.890$^\dag$ & 0.100 \\
    Abstract+Title & 0.313 & 0.103 & 0.388 & 0.206 \\
    \textit{all} (w/o authors)  & 0.322 & 0.133 & \textbf{0.401} & 0.211 \\
    \textit{all} & \textbf{0.348} & \textbf{0.170} & 0.695$^\dag$ & \textbf{0.239} \\
    \bottomrule
    \end{tabular}
\end{center}
\end{table}

The configurations that introduce a discount factor and integrate pseudo relevance feedback have an impact on precision (Table~\ref{tbl:cb-tfidf-conf-results}).
The precision of the recommendations increases for all data sets when the discount factor appears.
The same can be observed for the inclusion of pseudo relevance feedback, which provides the best configuration for the basic TFIDF approach.
The performance of the BM25 based recommendation strategy is similar to the PRF configurations.
The abstract meta-data benefits especially from the term weighting procedure, evidenced by the relative improvements over the PRF configuration being 16.6\% for the public groups data set.
Other types of meta-data do not seem to benefit from the term weighting.

\textbf{Hybrid Recommender System} The results from the best run configuration for the content-based, collaborative filtering and hybrid systems show that the latter technique performs the best (Table~\ref{tbl:hybrid-results}).
Again, out of the four test scenarios, the best results are achieved in the ``groups'' and ``publications'' ones.

\begin{table}[t!]
\caption{Comparison of the different configurations of the content-based approach using all available meta-data. Runs for the publications data set do not contain the author information.
}
\label{tbl:cb-tfidf-conf-results}
\begin{center}
    \begin{tabular}{ lrrrr }
    \toprule
    \textbf{Configuration} & \textbf{Groups} & \textbf{Venues} & \textbf{Pubs.} & \textbf{Libs.} \\ \midrule
    TFIDF & 0.348 & 0.170 & 0.401 & 0.239 \\ 
    NQ & 0.354 & \textbf{0.176} & 0.391 & \textbf{0.247} \\
    PRF & 0.358 & \textbf{0.176} & \textbf{0.415} & \textbf{0.247} \\ 
    BM25C & \textbf{0.361} & 0.157 & 0.402 & 0.244 \\
    \bottomrule
    \end{tabular}
\end{center}
\end{table}

\begin{table}[t!]
\caption{Comparison of the collaborative filtering (CF), the content-based (PRF) and the hybrid. The author information is excluded for the author publications data set.}
\label{tbl:hybrid-results}
\begin{center}
    \begin{tabular}{ lrrrr }
    \toprule
    \textbf{Algorithm} & \textbf{Groups} & \textbf{Venues} & \textbf{Pubs.} & \textbf{Libs.} \\ \midrule

    CF & 0.407 & 0.049 & 0.460 & 0.039 \\ 
    PRF & 0.358 & 0.176 & 0.415 & \textbf{0.247} \\ 
    Hybrid & \textbf{0.578} & \textbf{0.205} & \textbf{0.707} & 0.243 \\
    \bottomrule
    \end{tabular}
\end{center}
\end{table}

\begin{figure}[t!]
\begin{center}
\includegraphics[width=0.6\columnwidth]{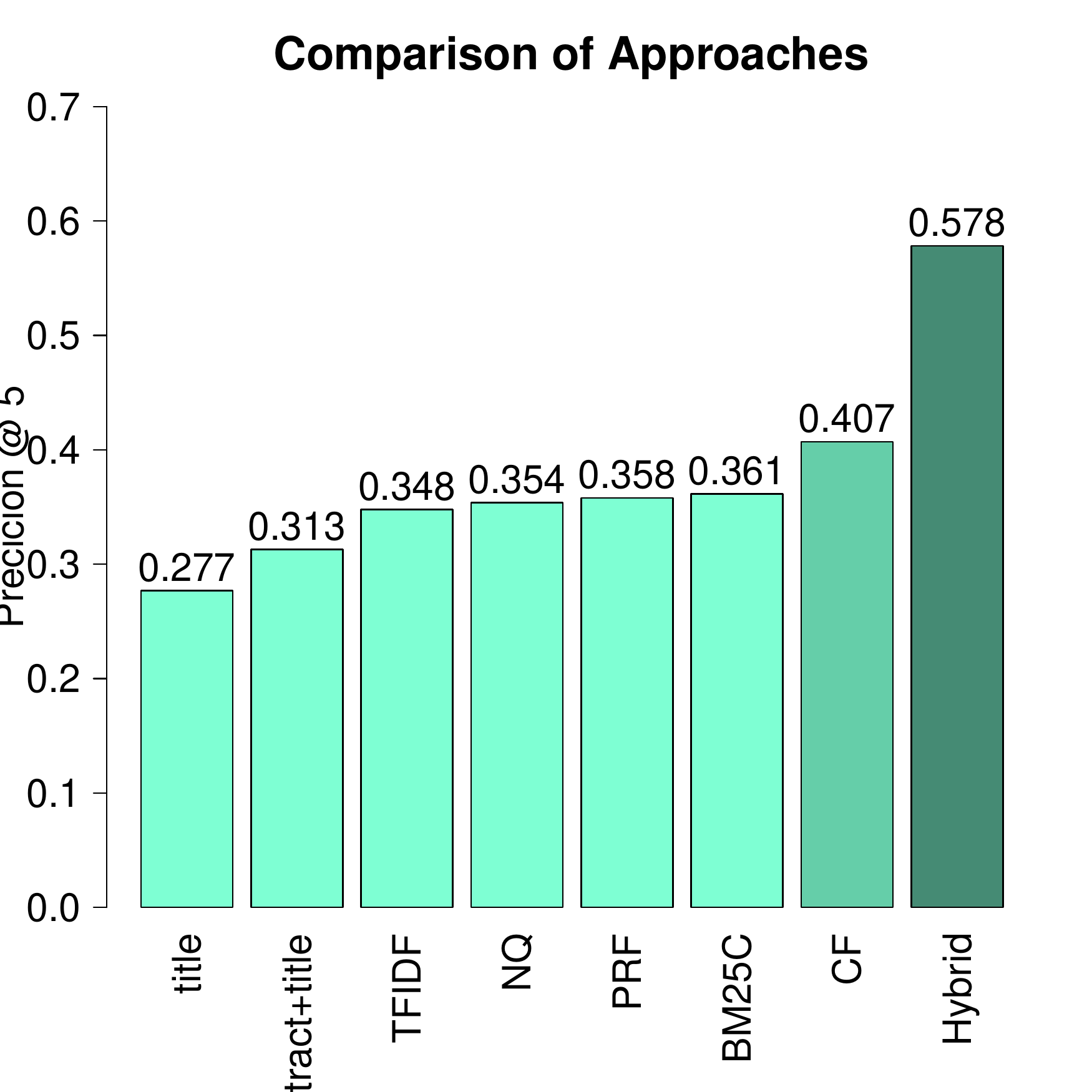}
\end{center}
\caption[Comparison of Approaches]{
Overview of the measured performance for all different approaches for the public groups data set.
Apparently the hybrid approach provides the best results.
}
\label{fig:approach-comparison}
\end{figure}

The various approaches are compared to each other for the public groups data sets (Figure~\ref{fig:approach-comparison}).
The same performance gain from the hybrid system can be observed for all other data sets as well, only for user libraries no improvement can be found.

\textbf{Sparseness} The evaluation tool calculates the results across all queries tested and across only queries where the model attempts to produce a result.
Some of the meta-data types are relatively sparse, thus the overall numbers might not reveal their potential for finding related articles.
The performance numbers for the collaborative filtering approach are held back due to data sparsity. 

\section{Discussion}


\begin{description}
    \item[Data-sets as ground truth] The first way to interpret the results is arguably more natural.
    Here the data sets can be seen as ground truth and the higher the performance numbers, the better the recommendation algorithm should be for the use-case represented by the data set.
    
    \item[Algorithms as ground truth] Another perspective would be to assume that the recommender provides the ground truth and the achieved performance indicates how well the data sets match the relatedness criteria embedded in the recommender approaches.
\end{description}

\subsection{Data-sets as Ground Truth}

\textbf{Collaborative Filtering} For many people a recommender system is synonymous with collaborative filtering.
These kind of systems often perform well, given enough training data.
In the four evaluations conducted, collaborative filtering produced excellent results in two scenarios, but failed in the other two.

Defining relatedness based on public groups is quite appealing, given that their contents have been hand-curated by Mendeley users in social, collaborative environments.  
Using Mendeley libraries to train a recommender that then predicts their contents well suggests that a similar kind of information is captured in both types of data set.  
Collaborative filtering also performs well in retrieving articles that have been written by the same authors (``publications'' scenario).  
This suggests that Mendeley users tend to read articles by the same authors and supports the use of the ``publications'' data set as a valid representation of an information retrieval strategy followed by users for finding related articles.

This algorithm, however, fails in the ``venues'' and ``user library'' settings.  In the case of ``venues'', this does not pose a particularly worrying problem given that content-based strategies perform so well.  
In the case of ``user libraries'', however, it is surprising to find that a collaborative filtering algorithm, trained on article co-occurrence in user libraries fails to perform well in predicting related articles, defined by co-occurrence in user libraries.  
Given that the type of relatedness captured by the training data should be comparable to that of the testing data, it is reasonable to assume that the algorithm failed since it did not have enough relevant co-occurrences in the training set.  
This scenario requires further investigation.

\textbf{Content-Based Filtering} For the content-based recommender, combining meta-data fields together improves the performance.
So did the integration of more sophisticated retrieval algorithms.
To further improve the performance, a deeper analysis of the individual meta-data values, such as named entity extraction, could be applied.

\textbf{Hybrid System} Generally the hybrid approach delivered the best results.
This can be seen as an indicator that meta-data and shared readership provide different types of evidence to indicate a relatedness relationship between articles.
Given the simple approach followed in combining the two sources of information, the results are even more pleasing.
Collaborative filtering would profit from a higher amount more users, who use the system. 
Whether such a system would approach the performance of the combined system is open for speculation.

\subsection{Algorithms as Ground Truth}

\textbf{Public Groups} 
Public groups are jointly used by multiple users and most of them are focused on specific topics.
The good performance of collaborative filtering suggests that the implicit agreement via user libraries is a good approximation for the explicit collaborative behaviour.

\textbf{Venues} The performance for the venues data set came as a surprise, because one would expect that users would follow the strategy of reading multiple publications from the same venue.
The results for the content-based recommender are worse than for the public groups, indicating that venues tend to not only focus on single topics, as apparently there is a wider variety of words being used.

\textbf{Author Publications} Both recommender approaches deliver the best results for the author publication data set.
This indicates that people tend to add multiple articles from the same authors to their collections.
Furthermore publications from the same authors appear to re-use the same words, which can be seen as a proxy for specific topics.
Closer inspection of the content-based filtering result indicates that articles from the same authors tend to be more similar when looking at the abstracts than when looking at titles alone in relation to the other data sets.

\textbf{User Libraries} The wide gap in performance between the ``groups'' and ``user library'' data sets suggest that users tend to organise their own libraries differently from those of the groups that they create.
It's important to note that since the data sets have been constrained to include only article sets of a cardinality between 10-20 articles, in terms of the ``user library'' set, this will mainly include researchers in their early stages.
These users apparently differ from the majority of users and they include a high variety of topics in their libraries, as indicated by the result from the content-based systems.

\section{Conclusion}

This work has been motivated by improving tools that support researchers in finding related and relevant research literature.
In doing so, four data sets have been constructed, each representing different information retrieval strategies for article relatedness.
Three recommendation system strategies were selected and evaluated against the four data sets.
For two of the data sets the collaborative filtering approach delivered better results than content-based filtering, but performed particularly badly in the other scenarios.
The best results were achieved by a combination of the two recommender systems, which consistently out-performed the two basic approaches.

There is a gap in recommender performance across the data sets.
The results indicate that the four data sets do indeed represent four distinct information retrieval strategies.
In making three of the four of them publicly available, this work contributes to the continued develop of recommendation technologies in the domain of scientific literature.

Improving the recommender algorithms, especially integrating the citation network and usage clicks, is a logical next step.
Furthermore the performance of the recommendation could be improved by integrating personalised recommendations into the system.

\section{Acknowledgments}

This work has been funded by the European Commission as part of the TEAM IAPP project (grant no. 251514) within the FP7 People Programme (Marie Curie) and as part of the EEXCESS project (grant no. 600601) within the European Union Seventh Framework Programme FP7/2007-2013.

\bibliographystyle{abbrv}
\bibliography{umap-2013}  

\end{document}